# The MICADO first light imager for the ELT: an overview of the RAM analysis and consequent design and maintenance strategy

Federico Biondi*[a], Yann Clénet[b], Patrick Caillier[c], Veronika Wimmer[c], Tristan Buey[b], Luis Neumeier[a], Florian Lang[a], Sebastian Rabien[a], Eckhard Sturm[a], Richard Davies[a]

[a]Max Planck Institut für Extraterrestrische Physik, Gießenbachstraße, 1, 85748 Garching bei München, Germany; [b]Observatoire de Paris à Meudon, 5 Pl. Jules Janssen, 92190 Meudon, France; [c]European Southern Observatory, Karl-Schwarzschild-Straße 2, 85748 Garching bei München, Germany.

## ABSTRACT

MICADO is a cryogenic near infrared Multi-AO Imaging Camera and Spectrometer developed for the first light operations at the ELT. It will operate in a "Standalone" configuration with a Single Conjugate Adaptive Optics module for a nominal period of two years. After this time, the system will be re-arranged in the "MICADO-MORFEO" configuration, being able to switch between the SCAO and a Multi Conjugate Adaptive Optics module in the later phase of the project. The lifetime requirement of minimum ten years, together with other demanding requisites about its availability and reliability triggered a meticulous FMECA analysis mainly focused on developing robust maintenance strategies. In this paper, we outline the assumptions and the boundaries of the MICADO RAM analysis, a collaborative effort involving the Max Planck Institute for the Extraterrestrial Physics, the Laboratoir d'Études Spatiales et d'Instrumentation en Astrophysique and the European Southern Observatory, starting from the input provided by all MICADO partners. We describe how RAM aspects drove some design choices as well as the selection and use of components. We report the preventive and predictive maintenance strategies, which we considered to minimize the risk of instrument downtime in the high cost operational context of the ELT.

**Keywords:** RAMS Analysis, Maintenance, ELT, MICADO.

## 1. INTRODUCTION AND INSTRUMENT DESCRIPTION

Multi-AO Imaging Camera for Deep Observations[1],[2],[3],[4],[5],[6],[7] (MICADO) is the Extremely Large Telescope[8] (ELT) first-light cryogenic imager and spectrometer. MICADO has been developed to work in the near infrared band, from 0.8 to 2.4 microns, mounting a 3 x 3 4k2 HgCdTe detector array[9][10]. It will provide diffraction limited images when assisted by adaptive optics correction. The available observing modes include classical imaging in low-resolution (4 mas/px, 50" x 50" FoV) and high-resolution (1.5 mas/px, 19" x 19" FoV), astrometric imaging with a differential precision of 50 µas and a calibration routine using an integrated mask[11], long slit spectroscopy with a spectral resolution of R ~ 20,000, and high contrast imaging[12],[13].

All the functionalities of MICADO are possible introducing, in a mixed warm and cryogenic system, a consistent number of motors and bearings, a solid system of sensors to constantly record physical and functional parameters, all orchestrated by a complex electronics and software system[14]. To guarantee a minimum period of ten years for the operational life of the instrument, and to minimize the downtime due to possible corrective maintenance activities, the consortium started an analysis of the components already during the Preliminary Design Phase, and delivered a complete FMECA and RAM analysis for the final design review of the instrument. The final goal of this exercise is the enhancement of reliability and availability, banally to maximize the number of observing nights of the instrument.

*biondi@mpe.mpg.de; phone +49 8930000-3283

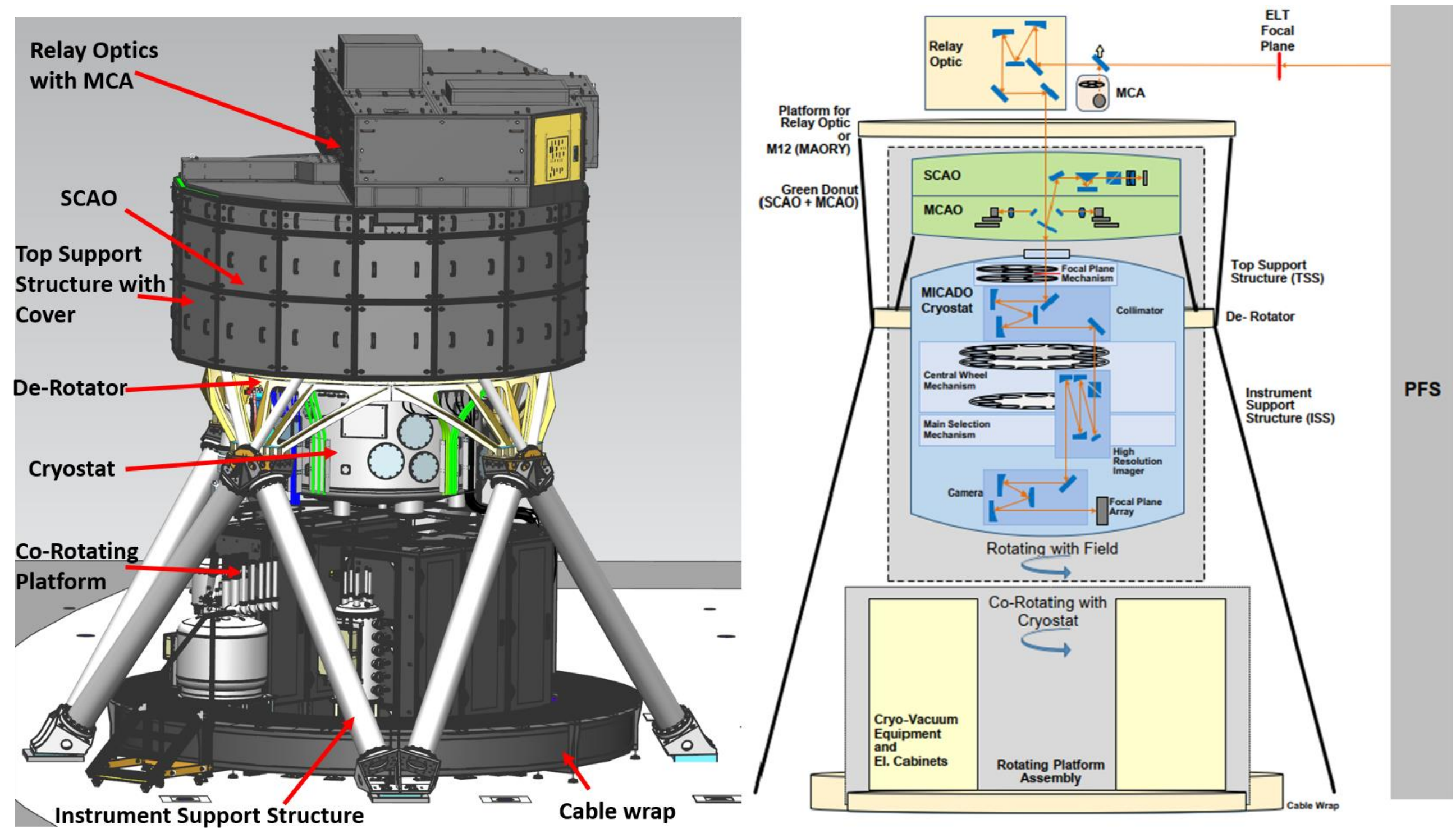

Figure 1. Rendering and functional design of MICADO. The access structure is omitted.

However, given the high cost per night of the ELT infrastructure, the goal of minimizing the downtime of the telescope represents also a mitigation of the financial loss of any downtime.

A more detailed description of MICADO helps understanding our approach to its RAM analysis, the results and the successive maintenance strategies. An effective way to depict the system is describing the main sub-systems following the path of light (Figure 1), highlighting the components that are relevant for the RAM analysis.

The light from the telescope enters the system through an open shutter in the Relay Optics[15],[16],[17][18] (RO). This is an opto-mechanical system with the scope to deliver the beam of the ELT minimizing the wavefront aberrations, the geometric distortion, the vignetting at the MICADO cold stop and the transmission losses. It also provide an intermediary pupil plane and it is used to absorb the I/F tolerances on both ELT and MICADO sides. RO has the capability to pick up the calibration light from the MICADO Calibration Assembly (MCA), with a mirror mounted on a rotational stage. Moreover three of the six RO mirrors have remote adjustment capabilities, used during the AIT.

The RO bench hosts the MCA[19],[20],[21], which in turn is composed by three sub-modules: a flat field and wavelength calibration unit, containing four gas lamps and a photodiode to monitor their status. The second module is a movable source calibration unit with VIS and NIR light sources, mounted on three linear stages and capable of patrolling the SCAO wavefront sensor field of view. The third module is an astrometric calibration unit including a pinhole mask mounted on a hexapod. RO and MCA are shown in Figure 2.

Under the RO bench, there is the Single Conjugate Adaptive Optics module[3],[22],[23],[24] (SCAO, Figure 3). The main sub-assembly hosted by its bench is the Natural Guide Star Wavefront sensor[25]. Besides the opto-mechanical components, this system includes the camera, and a number of motorized functions like piezo-mike actuators for mirror alignment, linear stages (i.e. field selector input mirror), linear stepper motors (i.e. rotation of field selector), an Atmospheric Dispersion Corrector, a K-mirror, an NCPA phase mask wheel, as well as angular and linear sensors. SCAO is equipped with a calibration channel, which includes a deformable mirror, again stages for mirror alignment, and a source selection module, which carries the fibers connected to IR and VIS fluxes. The calibration unit can be selected in the optical path using a deployment arm. A dichroic assembly, used to split the light into the $\lambda > 0.96$ µm (sent toward MICADO) and the $\lambda < 0.96$ µm (sent toward SCAO) parts, is allocated under the bench on a deployable stage.

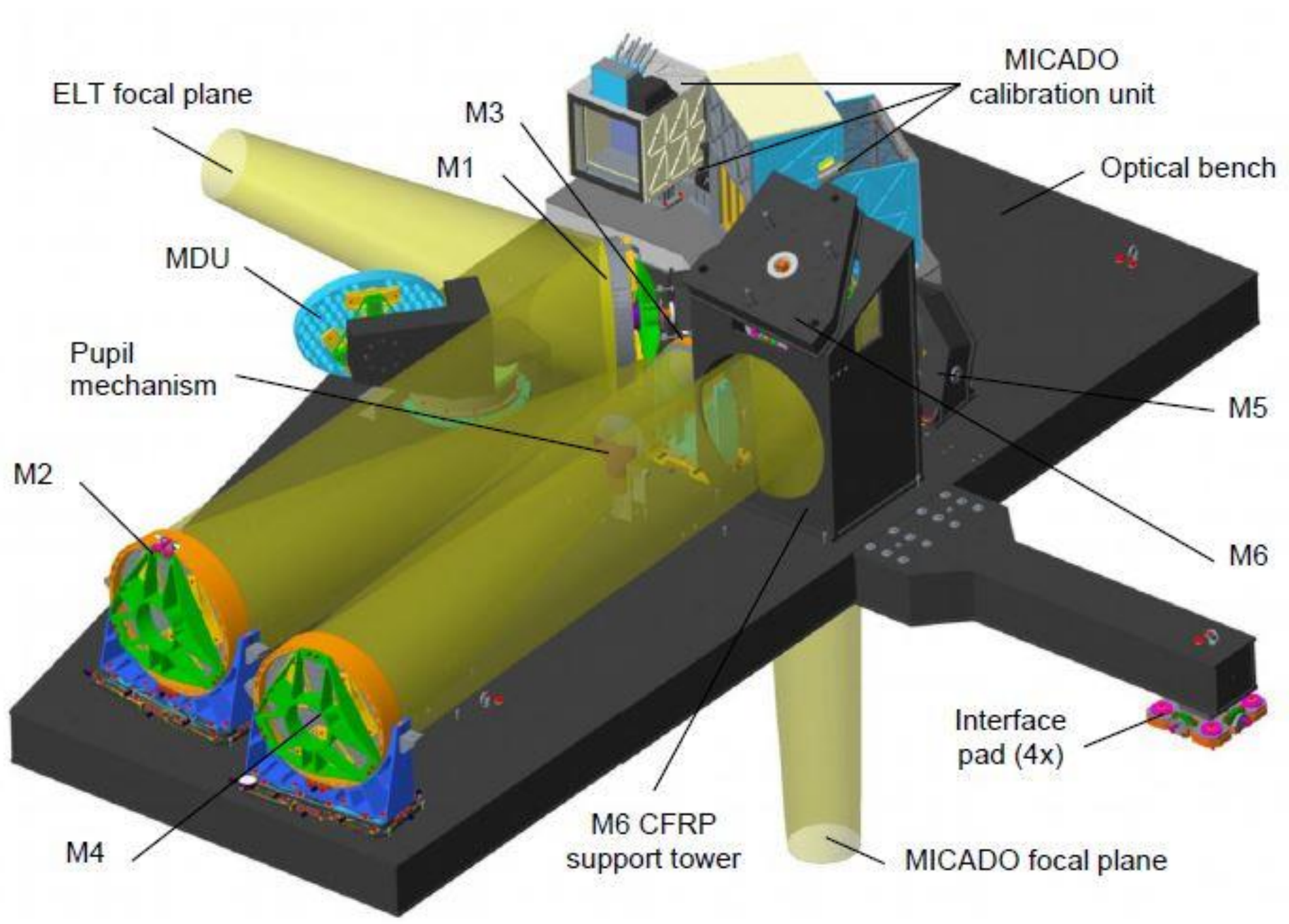

Figure 2. The Relay Optics hosting the Calibration Assembly.

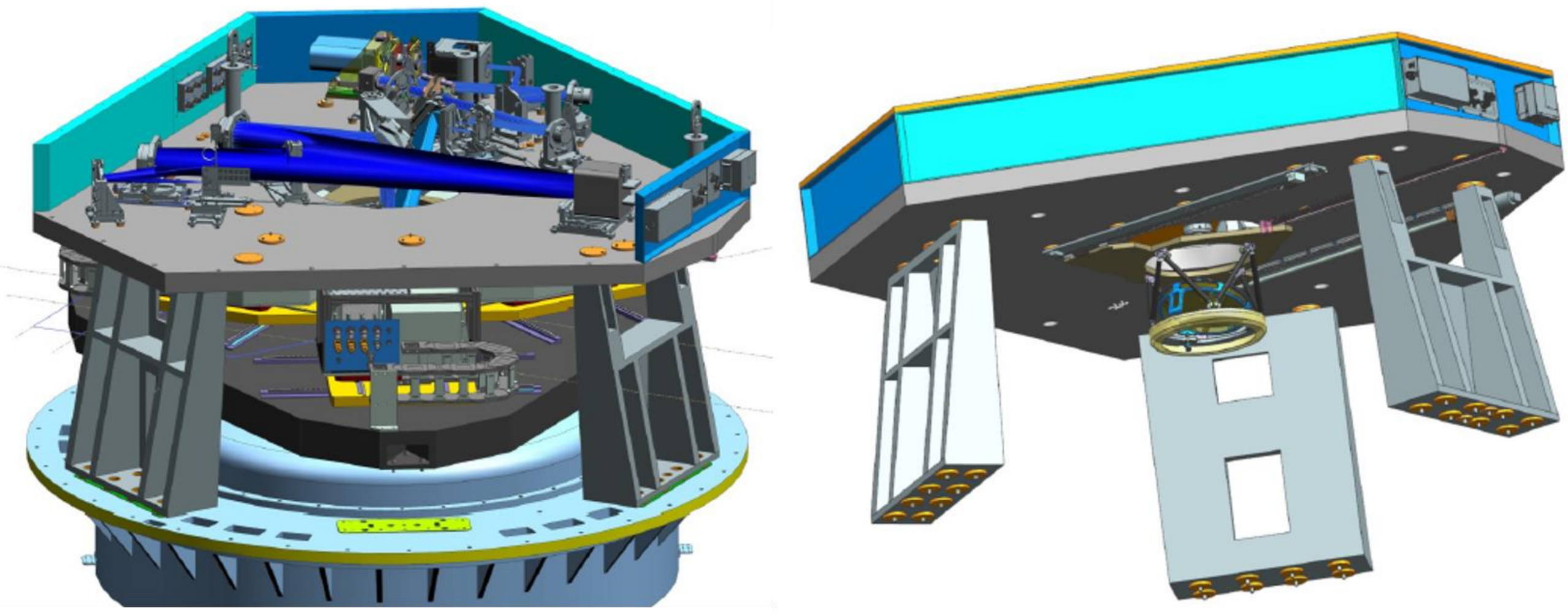

Figure 3. An impression of the complexity of the SCAO optical channels and a bottom view to locate the dichroic assembly.

After the dichroic, the light can enter the Cold System[26] (Figure 4), composed by the cryostat[27] (and its cryo-vacuum equipment, including also sensors for temperature, humidity, pressure, volume flow, filling level, and heaters), which is populated with the cold optics[28],[29], mechanisms, and the detector array[9],[10]. In particular the mechanisms are:

- Focal plane mechanism[30], which includes the "aperture wheel" and the "focal plane wheel".
- Central wheel mechanism[31],[32], which includes the two "filter wheels", a "pupil wheel" and an Atmospheric Dispersion Corrector.

- Main Selection mechanism[33],[34],[35], which allows switching between the high resolution imager, the low resolution imager, the spectrograph and a pupil re-imager.
- Detector positioning system[35], holding the detector and allowing its re-focussing capabilities during the AIT.

The nominal operational temperature for the systems inside the cryostat is 82 K.

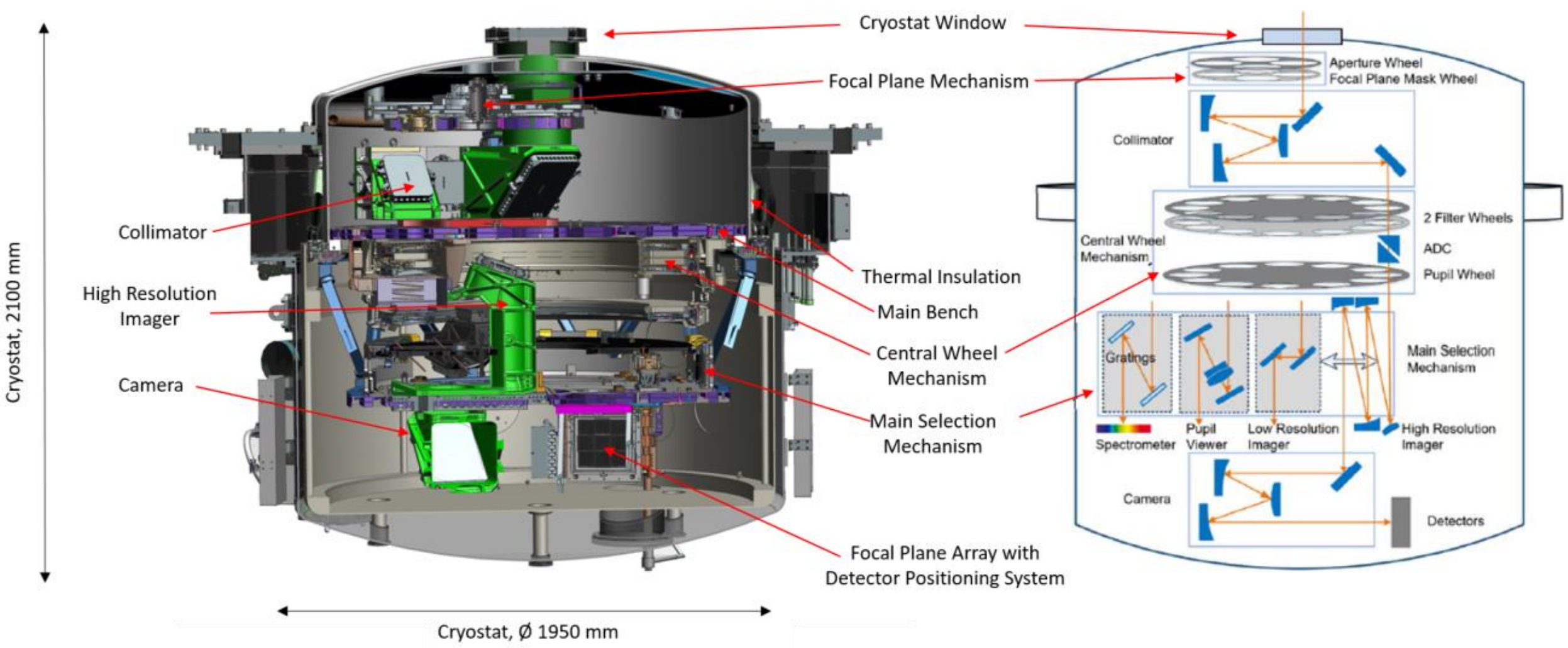

Figure 4. Section and functional design of MICADO cryostat.

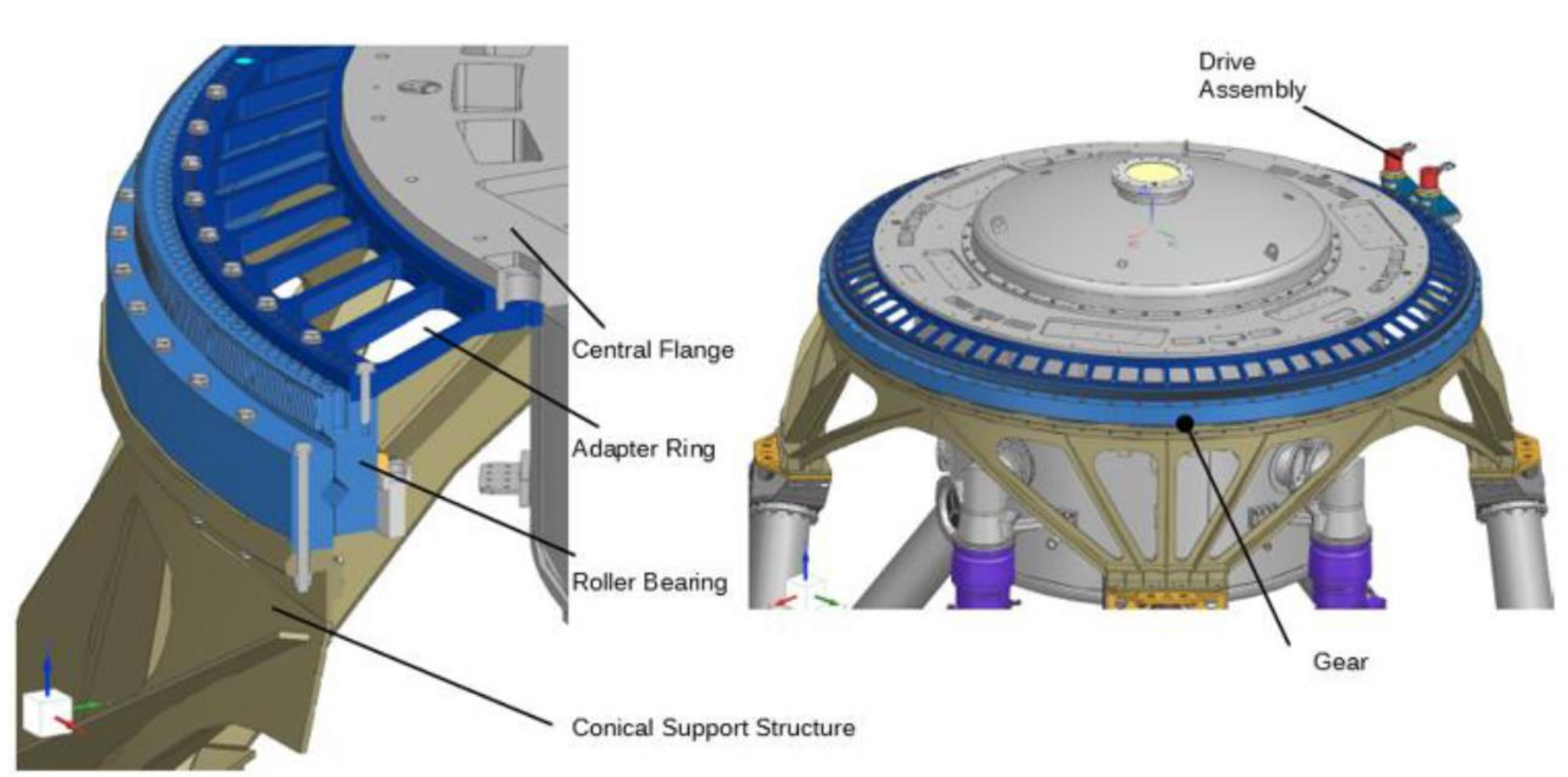

Figure 5. The Cryostat mounted on the De-rotator, which is supported by a conical interface with the Instrument Support Structure. On the left, a section of the De-rotator assembly.

The cryostat is mounted on the de-rotator (Figure 5), which is a roller bearing moved by two servo-motors, and allows to track the field of view or the pupil during an exposure.

These sub-systems are integrated on a support structure[37],[38], over a co-rotating platform[39] (Figure 6), which carries the electronics, the phase separator and the cooling system and which rotates the cable wrap following the de-rotator. Surrounding the whole system, there is a two-floor access structure.

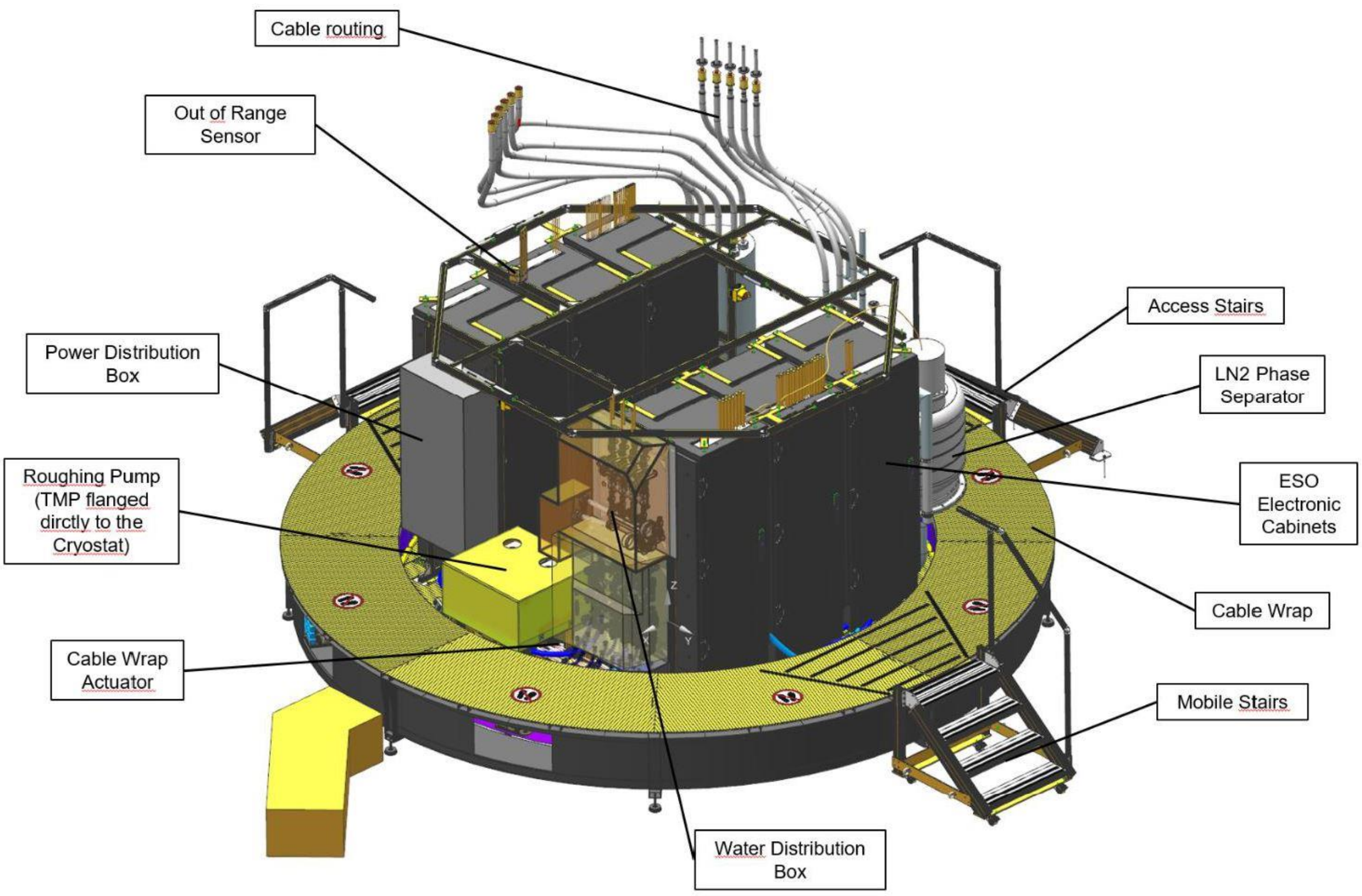

Figure 6. The Rotating platform surrounded by the cable wrap and carrying the electronics cabinets, the phase separator and the water distribution system.

The described configuration is called "Standalone" and will be installed on the Nasmyth Platform A (NAS-A) of the ELT for a nominal operational period of two years. Then MICADO will be dismounted from the NAS-A and re-arranged in the "M&M" configuration on the NAS-B, together with MORFEO[40], an instrument developed by a Consortium led by the Italian National Institute of Astrophysics, which will provide an additional MCAO capability. The Standalone to M&M reconfiguration consists in:

- De-commissioning of the RO, which is replaced by the last mirror of MORFEO (whose cover is visible in Figure 9, left panel).
- Re-configuration of the MICADO Calibration unit: its three sub-systems shall be re-arranged on a bench, which will be mounted onto the MORFEO volume; moreover, three additional calibration sources will be added.
- The "Low Order and Reference" module of MORFEO is mounted in the MICADO volume, just under the SCAO, replacing its mass dummy. The RAM budget of this sub-system is part of MORFEO RAM budget.
- Re-shuffling of the electronics cabinets.

For the MICADO RAM analysis, the applicable configuration is the Standalone configuration, since it is the most conservative in terms of components (presence of the Relay Optics) and operations (the SCAO is operational all the time).

We structured the rest of the paper as follow: in Section 2, we describe how the minimum lifetime for the sub-systems of MICADO can be reached, applying preventive maintenance actions. In Section 3, we discuss how we performed the FMECA analysis, what our “preventive strategy” is, and what are the design and predictive maintenance choices triggered by this exercise. Section 4 and 5 list the RAM analysis assumptions and caveats, and report the results compared with the requirements for ELT instrumentation. In Section 6, we show how we improved the maintainability of the instrument with appropriate access choices.

## 2. LIFETIME REQUIREMENT

The design of MICADO mainly comes from the heritage collected by institutions that developed comparable instruments for the biggest ground based telescopes as well as for space. The general guidelines applied for the design are the modularity, the standardization and the derating.

**Modularity**

MICADO has been designed following a modular approach, assuming the integration of sub-systems analyzed, manufactured, assembled and tested separately in advance. The modularity results in these two main advantages from a RAM point of view:

- Failure propagation can be stopped at the boundary of a module, when appropriate failure detection and isolation measures are implemented. In this way, we increase reliability, the maintainability and so the availability.
- Maintenance is simplified because line replacement is possible. High modularity of MICADO allows identifying Line Replaceable Units (LRU) during the design phase and providing spare parts and modules together with the delivery of the principal hardware to the telescope.

**Standardization**

Whenever possible, standard components, materials, procedures are included in the design. Standardization comes from the ESO heritage gained throughout previous projects. Main advantages are:

- The knowledge of components and associated statistics increase the reliability.
- The verification process and the analysis of the results are easier because of the presence of standardized procedures and pass-fail criteria, which are already available for past instruments or could be derived from component specifications.
- As long as components and devices underlie an official standard, it is easier to rely on different manufactures and vendors for the procurement.

**Derating**

We selected components preferably assuming to avoid operating them at the limits of their capabilities. This good practice applies in particular for assemblies operating inside the cryostat, since their maintenance is particularly time expensive.

General good practice *per se* cannot confirm the compliance with the requirement for ELT instruments about the total lifetime, which shall be longer than **ten years**: for MICADO this would include both the Standalone and the M&M phases. Another requirement about the lifetime asks explicitly to implement preventive maintenance and / or to plan overhauls whenever some part have a lifetime shorter than ten years. For MICADO, an overhaul is possible every three years. Moreover, as explained in Section 1, after two years of operations, MICADO shall be dismounted from NAS-A to be rebuilt on NAS-B with MORFEO: this phase represents an additional opportunity for extraordinary maintenance. In Table 1, we identified components with a lifetime shorter (or equal) than required, and we propose an intervention.

The ADC mechanism[41] is the only cryogenic subsystem on which an accelerated lifetime test has been performed. The result shows that after 7.5 years, the mechanism manifests issues reaching specific positions, while after 8.5 years it stops moving. The cause is a significant wear on the roller wheels in combination with the torque delivered by the stepper motor. Since the test shows that, before breaking, there is a hint of the performance degradation, it is possible to plan an overhaul to change the item.

Table 1. List of the items that require preventive maintenance to reach the required lifetime.

| Subsystem / Component | Lifetime analysis and preventive action |
|---|---|
| Vacuum Gauges | Maximum lifetime of 2 years, according to ESO experience. Since we implemented redundancy, the gauges will only be exchanged if a failure occurs. |
| MCA Cabinet | Exchange spectral lamp Neon and Xenon every 2 years, Argon every 4 years. Krypton lamp lifetime is 10 years. |
| ICS Cabinets | Exchange the cabinet fan every 4 years. |
| RP cable wrap hoses | According to the manufacturer, hoses need to be replaced every 5 years. |
| NGC II fans, dust filters and SFP+ modules | According to the ESO these parts need to be replaced every 5 years. |
| De-Rotator drive and gear | According to the manufacturer a lifetime of >10 years will be achieved with proper maintenance (lubrication). |
| RP central bearing | According to the manufacturer a lifetime of >10 years will be achieved with proper maintenance (lubrication). |
| RP Cable Wrap | According to the manufacturer a lifetime of >10 years can be achieved. We foresee a yearly inspection. |
| CWM Bearing Motors Sensors | Designed for a lifetime of 10 years, but might degrade faster than expected. Remote operation foreseen to check the status. |
| NGC II power supplies | According to the ESO the power supplies need to be replaced every 10 years. |
| Piezo Head modulation | The lifetime > 10 years requires preventing the local humidity to be above 85%: local humidity is constantly monitored and dry air is sent into the box when the humidity reaches a given threshold. |
| Field Selector DC motors | Lifetime is smaller than 10 years. We plan to change them at every overhaul maintenance operation. |
| Broadband lamp (in the co-rotating SCAO cabinet) | Lifetime is about 1 year so we plan to change it every year. |
| FS & PIL motor (PI L220) | Lubrication needed every year. Replacement every other overhaul. |

Please note that it is common to receive the data about the lifetime of a component in terms of "cycles" rather than in hours (see also Section 3.1). For example, for the Piezo-mikes of the SCAO, the lifetime is defined in 1,000,000,000 steps. Here an additional exercise should be done, projecting the expected usage during AIT and operation into steps count, based on previous experience: since the step size is 20 nm, we have more than 20 m adjustment for each screw. The estimation is being able to perform more than 1000 alignment procedures. With a maximum of 500 alignment procedures in the AIT phase, and less than 10 alignment procedure each year, the lifetime for these items would be 50 years.

## 3. THE FMECA TRIGGERED DESIGN AND MAINTENANCE CHOICES

Starting from the Preliminary Design Phase of the instrument, the MICADO consortium performed a complete Failure Mode, Effects, and Criticality Analysis (FMECA): following a bottom-up approach, we identified the components that are relevant for the next reliability and availability assessment (Section 5) and analyzed their possible failures in term of these parameters:

- Cause of the failure.
- Phase of the project / operational mode (i.e. AIV, Cool-down or Warm-up, operational state).
- Propagation of the failure from the "local" to the "end" effect.
- Means of failure detection.
- Severity of the failure.

In particular, the last parameter allows splitting the failures into three categories: the Severity 3 failures will cause a loss of science observation time and trigger a maintenance action that cannot be deferred. Severity 2 ones will cause a performance degradation of the science observation. All the others are Severity 1.

### 3.1 Impact on instrument design

An important parameter we considered, which implies a different way to carry out the FMECA, is whether a component is a "cold" or "warm" item, i.e. if it works inside or outside the cryostat. In particular, for "cold" components, we combined the frequency of the failure occurrence, its severity, and the capability of the system to detect it, by assign a number (1 to 10) to these parameters, and multiplying them. Then, we planned a "preventive strategy", i.e. a list of actions aimed to reduce the overall risk. The preventive strategy actualizes itself in a number of updated design choices. We identified three ways to proceed:

- Adding redundancy. This is an effort-saving strategy if applied to sensors, which are relatively not expensive, do not affect the mass budget or volume constraints, and have a minor impact on the control software.
  Redundancy has been introduced for sensors (which in turn should be cryo-qualified), and resolvers for motors, which should have the option to switch to step counting for positioning and use the indent in/out switches as extra control that the mechanism is in place.
- Following the prescription coming from other instrument heritage: the best example is the implementation of the detector sub-system by ESO.
- For more complex, critical, and expensive sub-assembly, the FMECA triggered a re-thinking of the component characteristics, starting from the nominal specification. The consortium typically carried out this exercise for cold bearings and motors, together with potential providers. It consisted in procuring slightly modified versions of already existent components built for space instrumentation or cryo-environments. Usually the input for the company are physical parameters as the:
  - Operational temperature and allowed temperature gradients.
  - Nominal loads and survival loads (i.e. during earthquakes).
  - Speed.
  - Expected Lifetime.
  - Limit in physical dimensions.

The analysis performed by the company results in a custom design that in some cases comes together with the lifetime of the item, even if expressed in "cycles" rather than in hours, as for the linear actuator for the DPS. The company provided also recommendations for a de-rated use of the motors, using the 50 % of the rated current (to avoid overheating of the wires) and the velocity 50 round / minute, corresponding to 1/3 of the full speed (to avoid bearing wear). The bearings, in turn, have been design to minimize frictions, using gold coated material, or stainless steal bearing rings with ceramic balls, depending on the sub-system.

Two examples of specific research for a custom solution made in house are:

- The magnetically coupled worm gear of the Focal Plane Mechanism[42] (Figure 7), to drive the wheels under clear room, vacuum or cryogenic conditions. The worms and a matching tooth system from soft magnetically material are coupled via magnetic forces without contacting each other. The magnetic field lines in the system are closed locally at the worms positions, keeping other areas free from fields. With surrounding the worms by magnets, the coupling fields between worm and driven tooth system are high.
- The Active Thermal Switch system[43],[44] enhances the cooling rate of mechanisms (FPM, MSM) by a factor of ~ 2. The research conducted at MPE was focused on the contact material, its shape and contact pressure.

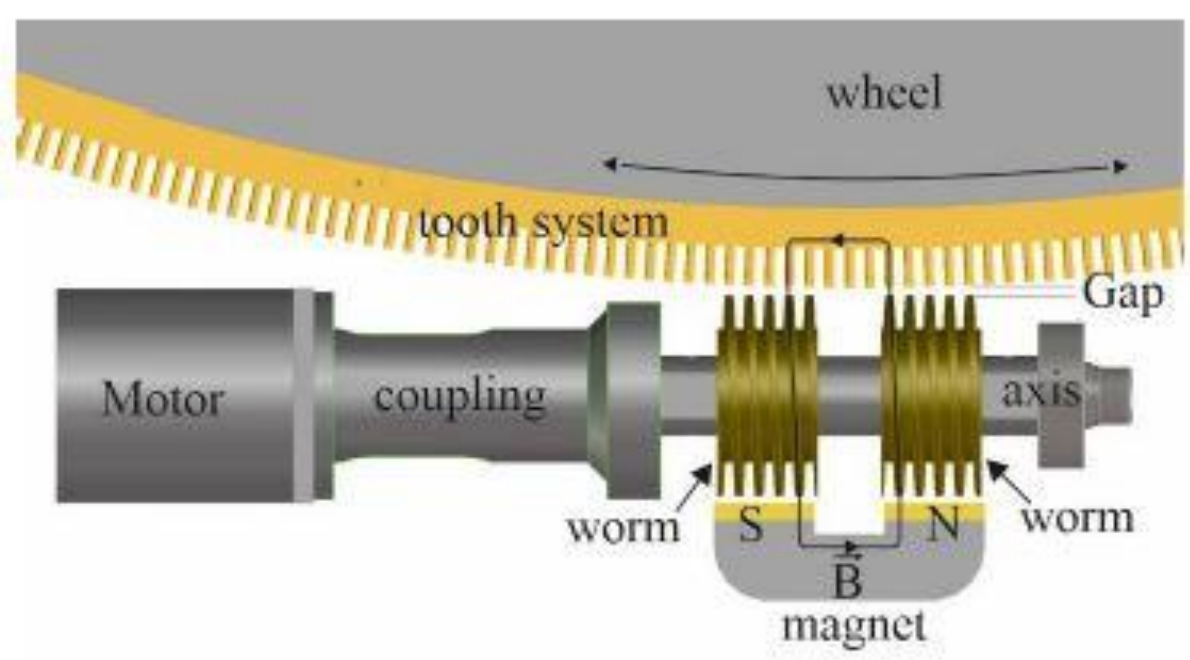

Figure 7. The magnetically coupled worm gear of the Focal Plane Mechanism.

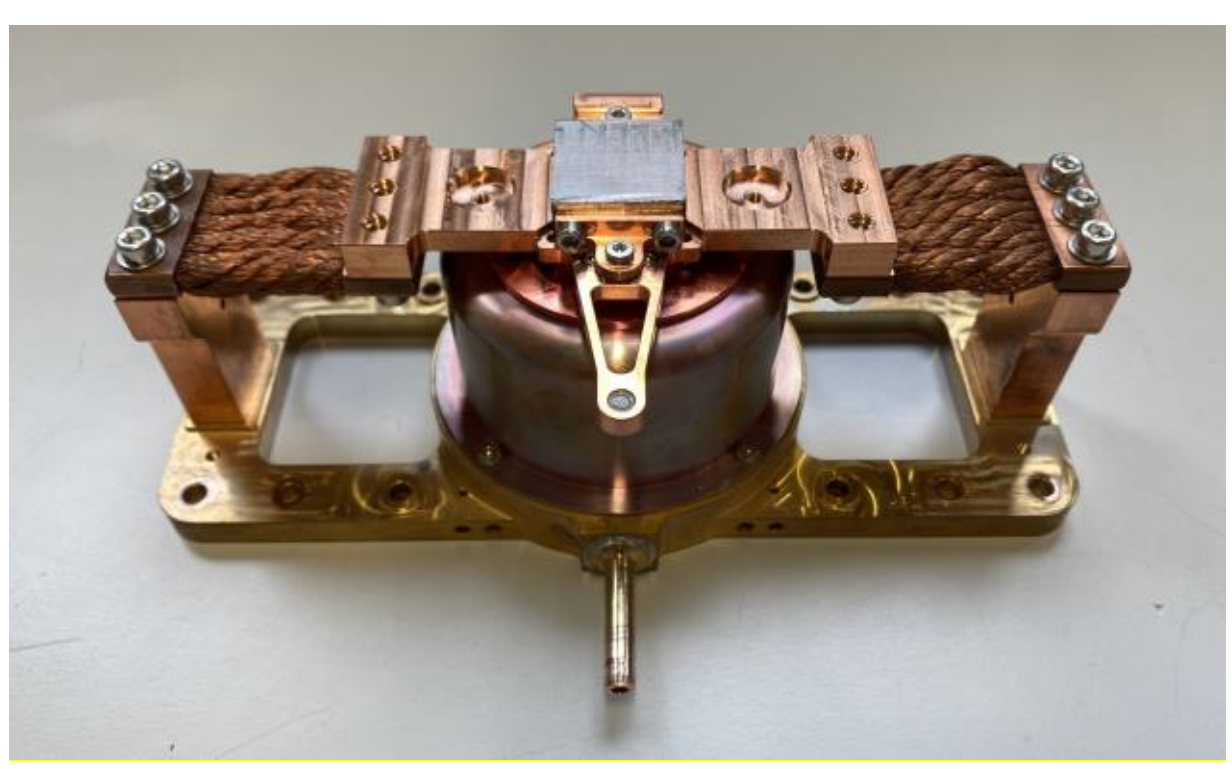
Figure 8. The Active Thermal Switch prototype tested in cryogenic environment.

Besides the design choices, the FMECA triggered also a specific test campaign at component level, aimed to certify mainly the electronic sub-assembly before the integration at system level.

Due to a lack of statistics about the reliability of the customized components, most of the cold items are not associated with any MTBF, making a formal approach to the subsequent reliability and availability assessment not possible. In Section 4, we explain how we approached the RAM for cold items in order to validate the requirements defined for ELT instruments.

### 3.2 Impact on maintenance choices

Excluding the components used only during the AIV, we were able to identify:

- 61 Severity 1
- 118 Severity 2
- 64 Severity 3

possible failures. In particular, 49 warm and 15 cold sub-assembly could trigger the Severity 3 failures. Within the warm ones, which are mostly electronic parts, replaceable in-situ within the next working day, we identified two items that require more effort to be repaired: the central bearing of the rotating platform and the energy chain of the cable wrap. These two sub-assemblies have been designed in collaboration with the manufacturer, and include a reasonable safety margin in terms of maximum lifetime. Preventive maintenance activities (i.e. inspection of the cable wrap, checking the motor for ball bearing noise, replace heavy duty wheels LRU if vibration or strong running noises occur) have already been detailed in the MICADO maintenance manual. Despite this, we anyway foresee a predictive analysis consisting in building a time statistics on two parameters: the torque of the bearing (through a measurement of the electrical current) and the measured load of the cable wrap. This approach enables disentangling the possible failure of the cable wrap from the one of the platform. In case of damage, to disentangle the effects of the motor itself and the ones of the bearing, we replace the motor-gear-drive unit with its LRU spare and check again the possible anomaly. Instead, to predict Severity 3 cold motor failures, we are investigating an alternative parameter. Cold motors are stepper motors, moved applying a constant current, independent by the load. As discussed in [45] and [46], a possible anomaly of the motor + bearing system could result in an increase of the torque and a consequent increase of the load angle. This parameter could be evaluated by the Beckhoff controller itself, provided that some additional motor parameters (i.e. EMF, coil inductance, Kp factor, Kl factor) are well known.

In the context of the predictive analysis, we will initiate the data acquisitions already in the AIT/V phases to arrive in an operational regime with a robust statistics that allows strengthening our predictive strategy.

# 4. ASSUMPTIONS FOR THE RELIABILITY AND AVAILABILITY ANALYSIS

In this Section, we list the assumptions on which we based our RAM analysis. These prescriptions are specified for MICADO as requirements, common to other ELT instrumentation.

- The temporal baseline for the analysis is the 10 years lifetime of the instrument.
- Assuming a fully equipped observatory composed by a suite of 7-8 instruments, each of them shall be in observation mode on average at least 40 nights per year. The night-time lasts typically 10 hours.
- Since MICADO will be a first light instrument for ELT, a stricter requirement for its time in observation mode is assumed: 80 nights per year, starting from a peak of 150 nights per year during the initial operation period, and going down to 40 nights towards the end of its lifetime.
- For each item, we indicated a duty cycle, typical values are:
    - 1.0 = always on
    - 0.5 = half time on (e.g. on during calibration and observation, off during day)
    - 0.026 = this is a “standard number” for cryo motors, which has been calculated from a typical observation scenario (corresponds to ~20 movements per mechanism of max. 2 min duration)

  We took into account the requirement about the operational time correcting the duty cycles by a factor 80/365 for component working only during an observation. These values are further corrected by a factor of 150/80 when calculating the burn-in period of two years.
- Static structural parts have not been considered for the RAM analysis.
- Even if the FMECA includes static electronics parts (harness, patch-panels, PCB-interface etc.), we have not considered them in the RAM analysis, bcause they do not have random failure mode.
- The components showing failure only during the cooldown or warm-up phases have not been considered in the RAM analysis.
- We considered the Standalone configuration applicable for the entire 10 years lifetime, even if the M&M configuration should be ready starting from the third year. This is a conservative approach in terms of RAM analysis, because it considers SCAO as the only AO facility for the entire lifetime.
- Another conservative assumption is to take into account the three extra light sources in the MCA, only used in the M&M configuration, for the whole ten years lifetime.
- Only Severity 2 and 3 failures have been considered relevant for the availability assessment.

As explained in Section 3, almost all the relevant components within the cryostat are customized solutions for MICADO. This prevents being able to receive a proper statistics about MTBF from the companies, and does not allow a formal approach to the reliability analysis. Our methodology consisted in assigning a unique MTBF of 400000 hours to every cold motorized functions, corresponding to a properly designed and derated product, but still not unrealistic. With this number, we were able to calculate a total MTBF for the system, which, for the cold instrument limited to Severity 3 failures is almost 86 times greater than the required one (Section 5), basically telling us that the assumption on the MTBF at component level could be relaxed.

## 5. RAM MAIN RESULTS

MICADO is requested to be compliant with a number of requirements, which describe an ideal reliability, availability and maintainability status. We report the most relevant of them in Table 2, Table 3, Table 4, for the reliability, availability, maintainability respectively, and we summarize results. Even if, formally, the reported results are not compliant with the requirements, it can still be considered that the instrument represents anyway a reliable and solid system.

Table 2. **Reliability** requirements and results expressed in Mean Time Between Failures, in hours.

| | Severity | Required MTBF [h] | Retrieved MTBF [h] |
|---|---|---|---|
| **Cold components / sub-assemblies** | **3** | 25920 | 2229592 |
| | **2** | 19440 | 22018 |
| | **1** | 12960 | 15947 |
| **Warm components / sub-assemblies** | **3** | 17280 | 3689 |
| | **2** | 12960 | 2499 |
| | **1** | 8640 | 4880 |

Table 3. **Availability** requirements in the operational and in the burn-in phase.

| Phase | Components / Sub-assemblies | Severity | Required Unavailability [%] | Retrieved Unavailability [%] |
|---|---|---|---|---|
| **Operational** | **Cold + Warm** | **2 + 3** | 0.84 | 2.98 |
| **Burn-in** | **Cold + Warm** | **2 + 3** | 2 | 3.10 |

Table 4. **Maintainability** requirements and results expressed in Mean Time To Repair, in hours.

| | Required MTTR [h] | Retrieved MTTR [h] |
|---|---|---|
| **Cold Components / Sub-assemblies** | 192 | 510.6 |
| **Warm Components / Sub-assemblies** | 3 | 9.7 |

### 5.1 Reliability

For ground-based instruments, the requirement for the reliability is basically linked to their availability, given the possibility to perform maintenance actions. The reliability requirement formally states that the proposed MTBF values are intended as a starting point to meet the availability requirement. With this interpretation in mind, we can assert that for warm components associated with Severity 1 failures, the non-conformance is totally solved considering their high maintainability, being reparable within maximum 2 hours (only the turbo molecular pump and the RPA heavy duty wheels take 8 hours). Moreover, as indicated in Section 4, for the availability assessment we shall consider only components associated with Severity 2 and 3. With an eye on the impact on the availability, we can say that the most relevant contribution comes from items associated with Severity 2 failures, which in turn still provide “partial” availability (causing only a degraded observation). In fact, considering only items associated with Severity 3 failures, the unavailability drops to 0.73 %.

### 5.2 Availability

The components that contribute significantly to the unavailability of the instrument are the cold ones. As described in Section 4, our approach to cold components and sub-assemblies was based on assigning them a unique and reasonable MTBF value to check *a posteriori* (i.e. after establish the availability parameter) if there is a margin to accept a lower MTBF. For this reason, looking to the detailed numerical outcome for the cold instrument availability has not solid bases. As described in Sections 3.1 and 3.2, to deal with cold components and downsize the failure risk, we already implemented preventive actions at design level with a careful choice of components, derating, and planning representative tests at components and sub-system level. Moreover, we are going to apply preventive and predictive analysis, and an evolving maintenance strategy starting from the experience of the AIT phase. Although not all these approaches have a direct impact on the retrieved unavailability values, they will contribute to the actual availability of the system.

### 5.3 Maintainability

In terms of formal compliance with the requirement for the cold components MTTR (192 h, as in Table 4), we should stress that the requested duration is not taking into account the 240 h needed to warm up and cool down again the cryostat. The warm-up and cooldown rates are in turn driven by the specifications for the thermal gradients on the detector.

For warm components instead, the main contributors to the MTTR are the field selector module and the dichroic sub-assembly in the SCAO. Please consider that, for these components, we considered the worst accessibility conditions and consequent worst MTTR case, corresponding to the Standalone configuration (even if this configuration should last for just two years). Moreover, we are considering the possibility to monitor motors parameters that can be related to a possible failure (as in Section 3.2).

## 6. MAINTAINABILITY / ACCESS OF MICADO

The total MTTR of the system depends on components' MTBF and MTTR, as well as from the number of spare parts. We discussed how we developed custom systems whenever needed to mitigate the risk of having un-reliable components in the design. In this section, we describe some design choices that have a direct impact on the MTTR. As described in Section 1, the design of MICADO is modular, and maximizes the presence of LRU, which can be exchanged in-situ by two technicians within eight working hours, without any other disassembly intervention. For each LRU, at least one spare unit will be available. If a quick corrective maintenance is not an option, the accessibility to the sub-units plays a fundamental role to mitigate the instrument downtime. This is particularly true for the interventions on part of the instrument, i.e. the systems placed into the cryostat that require dismounting part of MICADO, following the reverse order of the integration steps. With the purpose to facilitate the access to the instrument sections during the AIT and any maintenance routine, an Access Structure has been carefully developed, taking into account both the Standalone and the M&M configurations Figure 9.

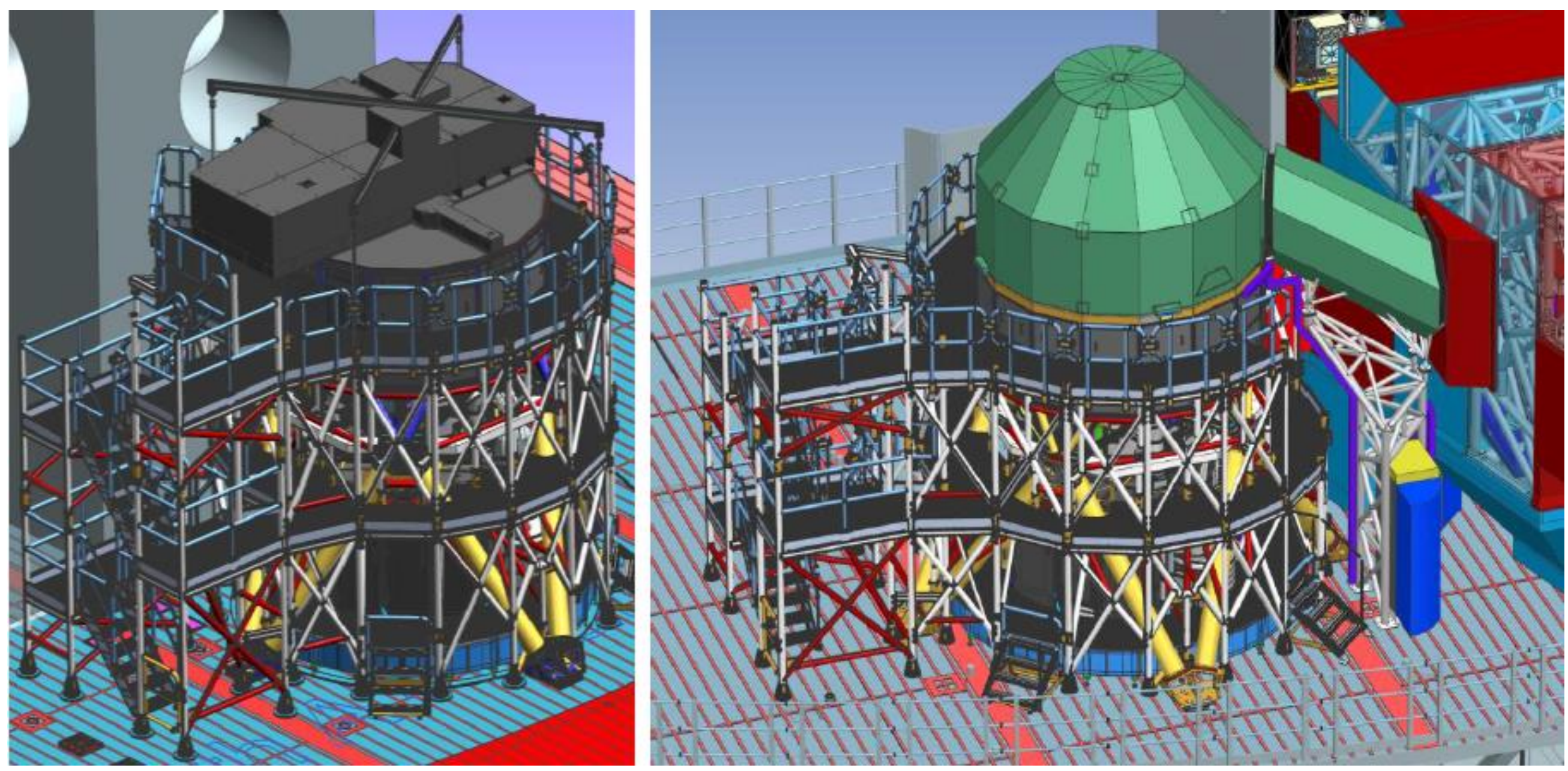

Figure 9. The Access structure in the MICADO Standalone (left) and M&M (right) configurations.

The Access Structure provides access to:

- “Ground floor” of MICADO for any intervention on the RPA, cable wrap, cabinets, phase separator and support structure bottom nodes.
- First level, at about 2.3 m, for the alignment of the support structure nodes, and for any AIT (i.e. reaching the crane and the sub-systems interfaces) and maintenance activity for the conical support structure, the de-rotator, and cryostat.
- Second level, at about 4.5 m, for any AIT and maintenance activity for the SCAO, RO (M12 in M&M configuration) and the top support structure.

To have accessibility to the entire 360° path, a “balcony” has been included to walk around the extruded volume of the RO. On that floor, an additional bridging ladder can be temporarily mounted to access the SCAO and partially the RO from the top; two beams shaped in a cross are used as a safety gantry to connect a harness (Figure 10).

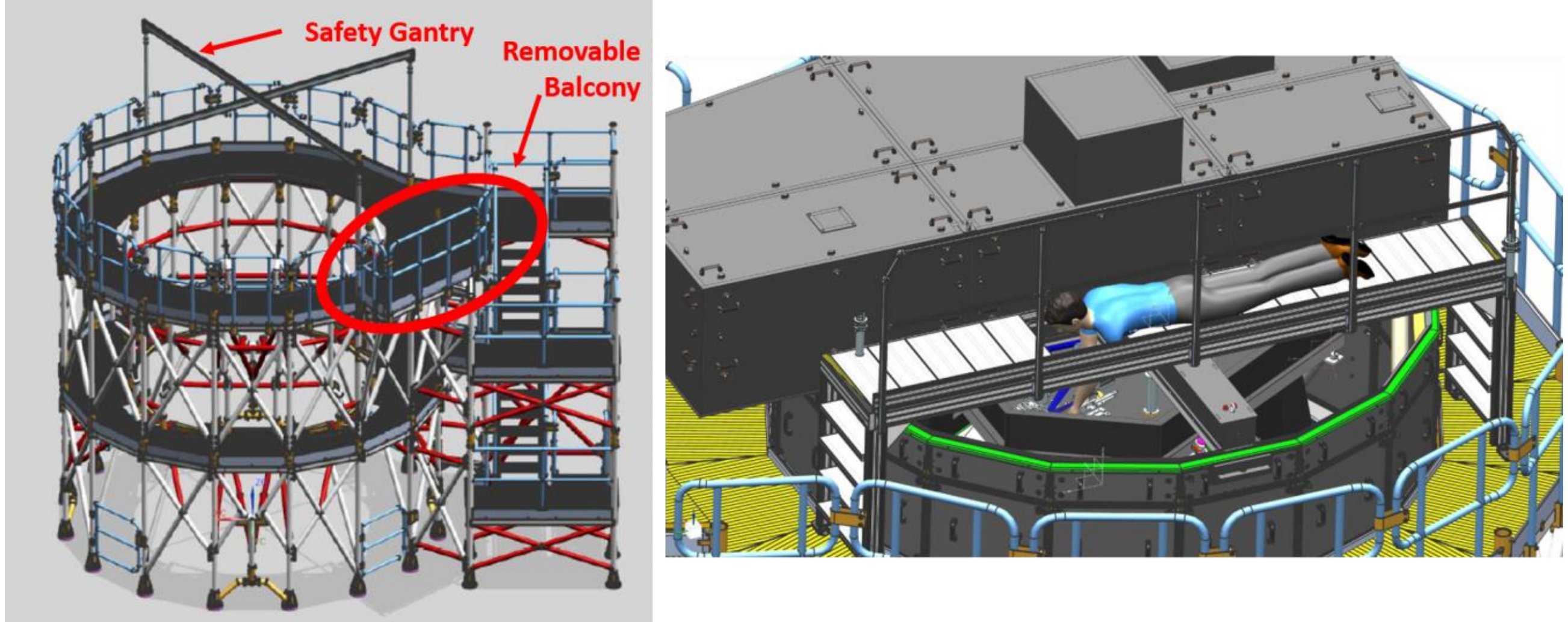

Figure 10. A solution to access easily the SCAO and the RO benches.

A peculiarity of the Access Structure is the possibility to accommodate additional scaffolding on the first floor for the maintenance of the LOR (the MORFEO AO module hosted in the MICADO volume). It allows also to dismount easily some of its segments (i.e. “angular slices” of the structure) to adapt the structure to the various need of the M&M configuration, in particular for specific integration and maintenance operation needed for MORFEO.

With this design, only some major interventions would require the disassembly of MICADO, reducing the time need for corrective maintenance and increasing the availability of the instrument.

## 7. CONCLUSIONS

We performed the FMECA and RAM assessment for MICADO, with a different approach for components operating inside and outside the cryostat. Since for most of the cold items, we miss the information about their MTBF, we implemented an optimized and customized cryo-design as a preventive strategy to reduce the risk of failures. For warm components, we developed a more formal FMECA and RAM analysis, identifying the most critical items, particularly in terms of MTTR. For cryogenic and RPA motors-bearing systems, we proposed a predictive maintenance strategy, based on the statistical analysis of motor parameters such as torque or load angle. This strategy will be tested and refined during the European AIT phase. We outlined our preventive maintenance plan, which applies to all the sub-systems with a lifetime smaller than the instrument's nominal ten-year life. Lastly, we described the improvements made on the Access Structure to ease the corrective maintenance of the sub-systems.

## ACKNOWLEDGEMENTS

MICADO is being designed and built under the leadership of the Max Planck Institute for Extraterrestrial Physics (MPE) by a consortium of partners in Finland, Germany, France, the Netherlands, Austria and Italy, together with ESO. In addition to MPE, the MICADO consortium consists of the Max Planck Institute for Astronomy (MPIA, Germany), the University Observatory Munich (USM, Germany), the Institute for Astrophysics in Göttingen (IAG, Germany), NOVA (Netherlands Research School for Astronomy, represented by the University of Groningen, the University of Leiden, and the NOVA optical/infrared instrumentation group based at ASTRON), INAF (Italy), CNRS/INSU (the National institute For Earth Science and Astronomy of the French National Centre for Scientific Research, represented by the LESIA Space and Astrophysics Instrumentation Research Laboratory at Paris Observatory, the GEPI Galaxies, Stars, Physics and Instrumentation Laboratory at Paris Observatory, the UTINAM Universe, Time-Frequency, Interfaces, Nanostructures, Atmosphere-Environment and Molecules Institute, the LCF Charles Fabry Laboratory at Institut d'Optique Graduate School, and the IP2I Institute of Physics of 2 Infinites), A* (an Austrian partnership represented by the University of Vienna, the University of Innsbruck, the University of Linz, the Johann Radon Institute for Computational and Applied Mathematics at the University of Linz, and the Austrian Academy of Sciences), FINCA (the Finnish Centre for Astronomy with ESO), and the University of Turku (Finland). This work has benefited from the support of 1) the French Programme d'Investissement d'Avenir through the project F-CELT ANR-21-ESRE-0008, 2) the project WOLF ANR-18-CE31-0018, 3) the CNRS 80 PRIME program, 4) the CNRS INSU IR budget, 5) the Action Spécifique Haute Résolution Angulaire (ASHRA) of CNRS/INSU co-funded by CNES, 6), the Observatoire de Paris and 7) the Ile de France region (DIM ACAV/ACAV+ and ORIGINES). USM is supported by the German Federal Ministry of Education and Research (BMBF) under the ErUM (Research of the Universe and Matter) grants 05A11WM1, 05A14WM1, 05A17WM1, 05A20WM1 and 05A23WM1.